\documentstyle[11pt,psfig,paspconf]{report}

\def\la{\mathrel{\hbox{\rlap{\hbox{\lower4pt\hbox{$\sim$}}}\hbox{$<$}}}}
\def\ga{\mathrel{\hbox{\rlap{\hbox{\lower4pt\hbox{$\sim$}}}\hbox{$>$}}}}

\newcommand{\dg}{\mbox{$^{\circ}$}}
\newcommand{\Hline}[1]{\mbox{H{\footnotesize {#1}}}}
\newcommand{\Hbeta}{\Hline{\mbox{$\beta$}}}
\newcommand{\HII}{\mbox {H\thinspace{\footnotesize II}}}
\newcommand{\etal}{{\it et~al.\/}}
\newcommand{\Msun}{\mbox{${\cal M}_\odot$}}
\newcommand{\pino}{\parindent=0mm}

\begin{document}

\title{THE CASE FOR SUBSTANTIAL DUST EXTINCTION AT {\em z} $\approx$ 3}
\author{\sc Gerhardt R. Meurer}
\affil{Department of Physics \& Astronomy, The Johns Hopkins University\\ 
(meurer@poutine.pha.jhu.edu)}

\abstract
Estimates of the total metal production rate ($\dot{\rho}_Z$) and star
formation rate at $z > 2$ are based on Lyman break systems
observed in the rest-frame ultraviolet (UV).  These observations are
very sensitive to dust obscuration.  Here I elucidate and refine the
Meurer \etal\ (1997) method for calculating UV obscuration, presenting
new relationships which accurately model the dust reprocessing of
radiation in local starbursts.  The median $\lambda = 1600$ \AA\
obscuration factor is $\sim 10$ at $z \approx 3$ which is shown to be
consistent with other constraints on the high-$z$ $\dot{\rho}_Z$.  Two
tests are proposed to further constrain $\dot{\rho}_Z$ at these redshifts.
\endabstract  

\section{Introduction} 

In their pioneering work, Madau \etal\ (1996; hereafter M96) use
Lyman break galaxies in the Hubble deep field (HDF), in conjunction
with other surveys, to evaluate the metal production rate
$\dot{\rho}_Z$ history of the universe.  They find that $\dot{\rho}_Z$
peaked at $z \approx 1$, while at $z \approx 3$ the universe was fairly
quiescent much like the present universe.  This result assumes dust
absorption is negligible.  However the high-$z$ observations, in the rest-frame
UV, are highly susceptible to the obscuring effects of dust.  Although
it is now recognized that absorption corrections are significant, the
amount of absorption is still under debate.  Following from UV
observations of local starbursts (Meurer \etal\ 1995, hereafter
paper-1) we have argued that the the correction factor to
$\dot{\rho}_Z$ is considerable (roughly 15; Meurer \etal, 1997;
hereafter paper-2), while others (e.g.\ Madau 1997, hereafter M97;
Pettini \etal\ 1997, hereafter P97) prefer fairly modest, $\sim$
factor of three, corrections.  As pointed out by M97, what is at stake
is more than just the amount of dust in the early universe.  If the
absorption corrections to $\dot{\rho}_Z$ are low, then hierarchical
galaxy formation models are favored, while large corrections at
high-$z$ can push back the peak $\dot{\rho}_Z$, favoring monolithic
collapse models.  Here I take the opportunity to elucidate and refine
our technique for estimating the UV attenuation due to dust,
reapplying it to high-$z$ galaxies.  Further constraints and
refinements to the high-$z$ star formation rate are discussed.
Finally, I present tests to further constrain the high-$z$ $\dot{\rho}_Z$.
Throughout this paper I adopt $H_0 = 50\, {\rm km\, s^{-1}\,
Mpc^{-1}}$ and $q_0 = 0.5$.

\section{Technique}

Before proceeding, consider the basic UV absorption and scattering
effects of dust. When observing individual stars through small
apertures, scattering removes light from the line of sight.  This is
the geometry used to derive traditional extinction curves. The
radiative transfer effects are different when observing galaxies,
because firstly, if they are observed through sufficiently large
apertures the forward scattered light is recovered, and secondly, the
distribution of the dust is important, with clumpy dust screen and mixed
and mixed dust geometries producing more absorption per unit
reddening than homogeneous foreground screens.  The net absorption as a
function of wavelength normalized to the optical reddening is the
``obscuration curve''.  Fortunately dust is not a sink for photons;
the absorbed radiation is re-emitted thermally in the far-infrared
(FIR).  For irradiation by young stellar populations, the dust heating
is dominated by UV photons.  Hence the ratio of FIR to UV fluxes is an
indication of net absorption.

\begin{figure}[t]
\centerline{\hbox{\psfig{figure=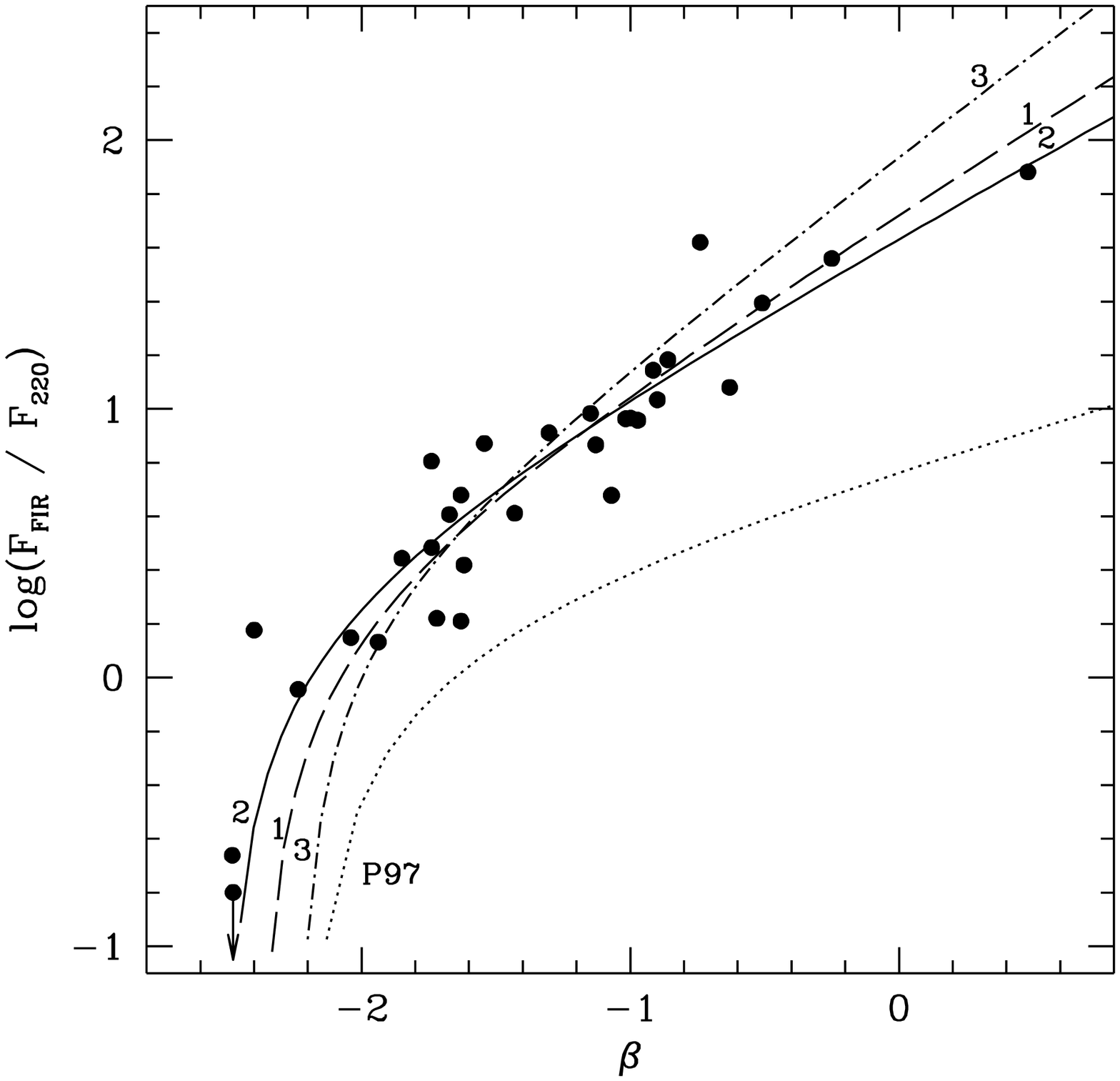,width=7cm}
\psfig{figure=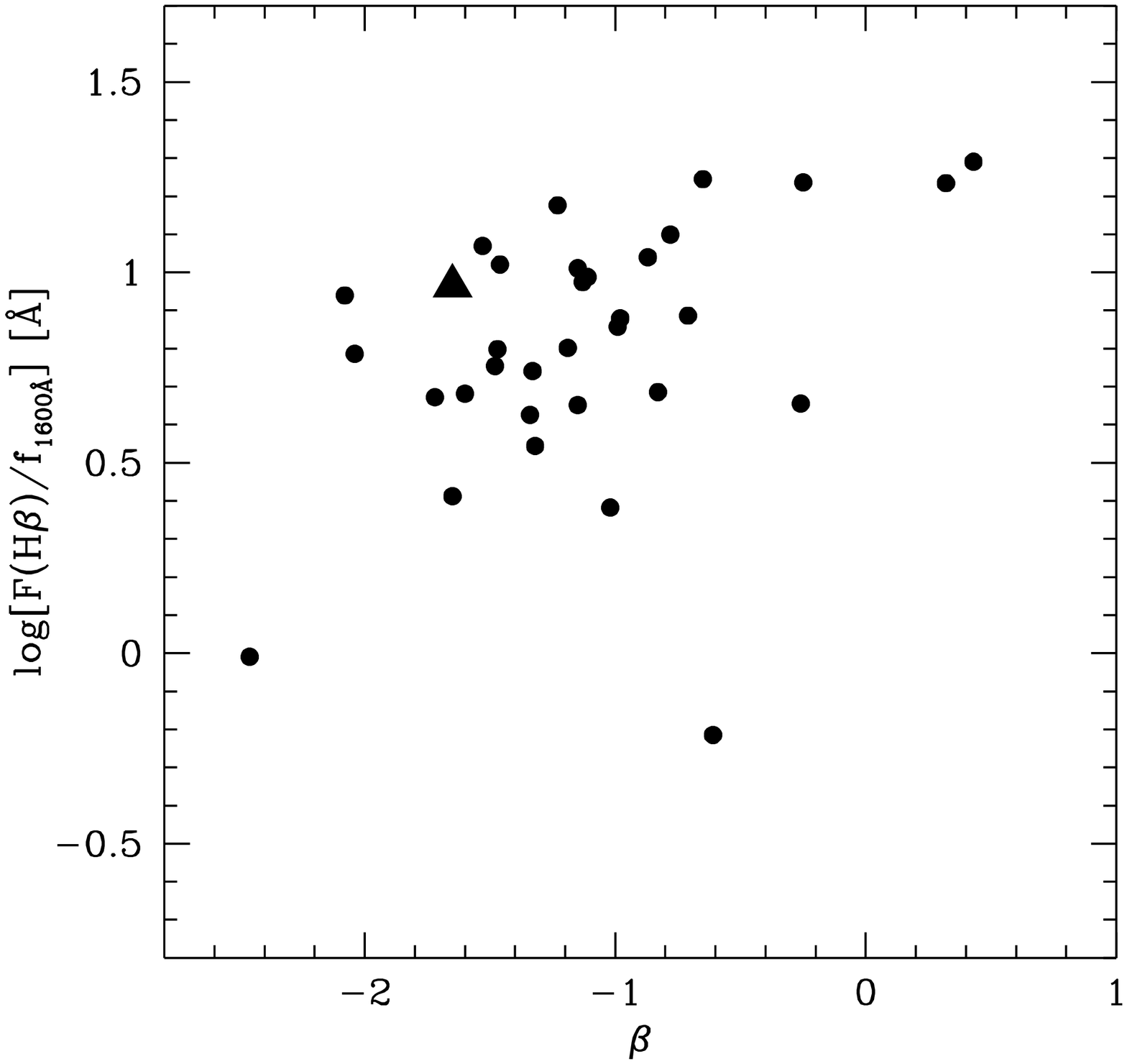,width=7cm}}} {\footnotesize 
{\bf Figure 1 (left).} Empirical UV absorption factor compared to UV
color $\beta$.  The fluxes are measured in the FIR by IRAS and in the
UV ($\lambda \approx 2300$\AA) by HST or IUE.  The numbered lines
correspond to the three model fits discussed in the text.  The dotted
line shows the proposed reprocessing model of P97.  \\
{\bf Figure 2 (right).} Raw \Hbeta\ fluxes normalized by UV flux
density plotted against $\beta$.  Circles show local starburst
galaxies (Calzetti 1997), while the triangle corresponds to the Lyman
break galaxy 0201-C6 (P97).}
{\small 
\begin{center}
{\bf Table 1} \\[2mm]
\begin{tabular}{|l | c c c c c |} \hline\hline
Model                       &      1 &             2 &      3 & paper-2        & P97 \\
&&&&& \\[-3mm] \hline
$\Delta t$ [Myr]            &     80 &            20 & \ldots &    10  & \ldots \\
Obscuration law$^a$         &    C94 & 0.8C94+0.2K94 &    C94 &    C94 &    SMC \\
$R$                         &    3.1 &           3.1 &    4.8 &    3.1 &    3.1 \\
$\beta_0$                   & --2.33 &        --2.45 & --2.20 & --2.53 & --2.13 \\
$\langle E(B-V)\rangle^b$   &   0.28 &          0.27 &   0.23 &   0.33 &   0.09 \\
$\langle A_{1620}\rangle^b$ [mag] &   2.41 &          2.41 &   2.51 &   2.79 &   1.07 \\[-3mm] 
&&&&& \\ \hline
\end{tabular} 
\end{center}
{\small {\bf Note: } \\
$^a$ C94 = Calzetti \etal\ (1994); K94 = Kinney \etal\ (1994).  \\
$^b$ Evaluated at median $\beta = -1.1$ (paper-2).   }
}
\end{figure}

Our analysis is based on the remarkable similarity of Lyman break
systems to local starburst galaxies.  Figure 1, adapted from
papers-1,2 demonstrates that, for local starbursts, UV absorption is
correlated with UV spectral slope $\beta$ ($f_\lambda \propto
\lambda^\beta$).  This gives us a powerful method of estimating
obscuration from UV colors.  Figure~1 shows our latest foreground
screen model fits to this relationship.  We assume that it is due to the
reddening of a single intrinsic spectrum, for which we adopt the
constant star formation rate models of Leitherer \&\ Heckman (1995;
Salpeter IMF with upper mass limit of 100\Msun) or power law spectra
models (to extend the range of intrinsic color $\beta_0$).  For
obscuration curves we choose those derived by Calzetti and collaborators
(Kinney \etal\ 1994; Calzetti \etal\ 1994; Calzetti, 1997) from
starburst spectra with varying amounts of \HII\ reddening.   Three
fits are shown.  Model 1 employs the Calzetti \etal\ (1994) curve,
assuming $R \equiv A_V/E(B-V) = 3.1$, and is fit to find the best
burst duration $\Delta t$.  Model 2 is the best fit at fixed $\Delta t
= 20$ Myr for obscuration curves varying linearly between that of
Calzetti \etal\ (1994) and Kinney \etal\ (1994), again with $R = 3.1$.
Model 3 is the best fitting power law spectrum for the Calzetti \etal\
(1994) law with $R = 4.8$ as determined by Calzetti (1997).  The
parameters of the three fits are given in Table~1.  I also give the
parameters adopted in paper-2, and those adopted by P97 (M97 adopt a
similar prescription).  The latter does not well model the dust
reprocessing of UV radiation observed in starbursts.  

Table 1 shows the implications of the models by listing the implied
obscuration at $\lambda = 1620$\AA, for an observed $\beta = -1.1$
(the rest $\lambda$ and median color of the HDF Lyman break systems;
M96; paper-2).  Our new fits all yield $\langle A_{1620}\rangle = 2.46
\pm 0.05$ mag.  The fits are nearly model independent since $F_{\rm
FIR}/F_{\rm UV}$ is the empirical obscuration factor, modulo a
correction (of order unity) to $\lambda = 1620$\AA.  The scatter about
the fits, amounting to a factor of 1.6, is the limiting uncertainty of
the calibration.  Starting with the luminosity function corrected
$\dot{\rho}_Z$ estimates of Madau (1996), our models imply corrected
values of $\dot{\rho}_Z = 6.4,2.1 \times 10^{-3}\, {\rm \Msun\,
Mpc^{-3}\, yr^{-1}}$ for the $U$ and $B$ band dropouts at $z = 2.75,
4.0$ respectively.  These estimates are still lower
limits because of the unknown contribution from galaxies completely
obscured in the UV by dust.  Note that this method can not be
blindly applied to $\dot{\rho}_Z$ measurements for $z < 2$, because
they are drawn from samples not necessarily dominated by starbursts.

\section{Other constraints on high-$z$ star formation, and further refinements}

\noindent
{\em $\bullet$ Extension to rest frame optical fluxes}.  Sawicki \&\ Yee
(1997; and this volume) fit the $VIJHK$ SEDs of the $z >
2$ galaxies in the HDF with population synthesis models, finding these
galaxies to have $E(B-V) \approx 0.2 - 0.3$, and 
$\Delta t \leq 0.2$ Gyr, consistent with our models (Table 1).

\noindent
{\em $\bullet$ FIR-mm background}.  Contrary to claims by M97 and
P97, the FIR-mm background flux measurements of Puget \etal\
(1996, and others) allow for substantial obscured high-$z$ star
formation (e.g.\ Guiderdoni \etal\ 1997).  If the dust temperature
$\approx 50\dg$K, as is the case for typical starbursts, then
$\dot{\rho}_Z \leq 0.056$ \Msun\ Mpc$^{-3}$ yr$^{-1}$ for $z \geq 2.5$
(Burigana \etal\ 1997).

\noindent
{\em $\bullet$ Metal content of the universe}.  M97 notes that
monolithic collapse models greatly overpredict the metal content of
the local universe and the metallicity of damped Ly$\alpha$ systems at
$z > 2$.  However, these estimates should be considered lower limits
since they do not include metals in the inter-galactic medium (IGM)
where much of the metals produced by starbursts are likely to be
ejected.  Mushotzky \&\ Loewenstein (1997)
estimate the metal content of the IGM (using ASCA observations) to be
2-5 times that in stars and the ISM (M96) and suggest
that the star formation that enriched the IGM occurred either at $z >
1$, or was dust enshrouded.  Indeed, the corrected high-$z$
$\dot{\rho}_Z$ agree well with their
predictions for $z = 1-6$.

\noindent
{\em $\bullet$ Low mass stars}.  At this conference, Madau suggested
that monolithic collapse models overpredict the near infrared (1-2
$\mu$m) luminosity density for $z < 1$. This assumes a universal
Salpeter IMF slope down to $\la 1$ \Msun.  However, there is some
evidence that the IMF in starbursts may be biased against low-mass
stars (Rieke \etal, 1993).  Furthermore, the discrepancy is only $\sim
0.3$ dex, which may partially be accounted for by systematic
uncertainties in the modeling.

\noindent
{\em $\bullet$ Other refinements}.  Two other refinements to our technique are
suggested by the work of Dickinson (this conference; also noted by
M97, P97) -- galaxy by galaxy absorption corrections, and better
modelling of the calibration between broadband colors and spectroscopic
$\beta$.  We will address these elsewhere, but the good agreement
between our results and those of Sawicki \&\ Yee suggests that these
effects are not large.

\section{Future constraints}

Our obscuration estimates are based on the premise that the Lyman
break systems are like local starbursts. They are overestimated if the
high-$z$ sample is significantly contaminated by intrinsically redder,
non-ionizing populations.  Two tests of the starburst
hypothesis are feasible:

\noindent
{\em $\bullet$ Emission lines.}  Detection experiments for recombination
lines in high-$z$ galaxies would directly determine whether they are
dominated by ionizing populations.  P97 note that
\Hbeta\ has been detected in three Lyman break galaxies and report
$L_{\rm H\beta}$ for one of them, 0201-C6.  In Fig.~2 I plot this flux
normalized by its UV flux (kindly made available by M.\
Dickinson) with data from Calzetti (1997) showing that 0201-C6 is
similar to local starbursts.  More observations, particularly of the
reddest systems, are needed to determine the degree of contamination by
non-starbursts. 

\noindent
{\em $\bullet$ Radio continuum.}  If obscured starbursts emit
primarily in the FIR, then we may hope to
detect them in radio continuum via the radio-FIR
correlation. Fomalont's \etal\ (1996) deep 8.4GHz image of the HDF has
a detection threshold of 12 $\mu$Jy corresponding to $L_{\rm bol}
\approx 10^{12.8}\, L_\odot$ for $z = 3$ and an assumed continuum
index $\alpha = 0.7$ ($f_\nu \propto \nu^\alpha$).  While none of the
galaxies detected by Fomalont \etal\ are Lyman break systems, only one
of the HDF $U$ dropouts has $L_{\rm bol}$ large enough that it could
(barely) have been detected.  Deeper radio imaging (perhaps at lower
frequencies) should better constrain the FIR emission of high-$z$
starbursts.

\section{Summary}

Our view of cosmological evolution is obscured by dust, particularly
at $z > 2$ where the observations are in the rest-frame UV.  The
method presented here of estimating the UV dust obscuration factor
accurately models the dust reprocessing of radiation in local
starburst galaxies.  With the assumption that Lyman break galaxies are
like local starbursts I estimate a median correction factor of $\sim
10$ to $\dot{\rho}_Z$ at $z \sim 3$.  The corrected $\dot{\rho}_Z$
estimates are consistent with other constraints on the high-$z$ star
formation rate.  Two tests of these results are feasible with existing
technology. Observations of rest frame Balmer emission provide a test
of the starburst assumption, while deep radio imaging can provide an
independent estimate of the reprocessed UV emission.

\medskip\noindent {\em Acknowledgements.}  This is a preliminary
version of a paper I am writing with Daniela Calzetti and Tim Heckman.
I thank them for their insights and for allowing me to publish this
alone.  I thank Mark Dickinson, Andy Connolly, Piero Madau, Max
Pettini, and Marcin Sawicki for useful discussions, and Annette
Ferguson for reading a draft of this paper.

\baselineskip=0.4cm \pino

{\footnotesize
\references
\reference Burigana, C., Danese, L., De Zotti, G., Franceschini, A.,
Mazzei, P., \&\ Toffolatti, L. 1997, \mnras, 287, L17 
\reference Calzetti, D. 1997, to appear in ``The Ultraviolet Universe
at Low and High Redshift: Probing the Progress of Galaxy Evolution'',
W.H. Waller, M.N. Fanelli, J.E. Hollis, \& A.C. Danks, eds. (astro-ph/9706121)
\reference Fomalont, E.B., Kellermann, K.I., Richards, E.A.,
Windhorst, R.A., \&\ Partridge, R.B. 1997, \apjl, 75, L5
\reference Guiderdoni, B., Bouchet, F.R., Puget, J.-L., Lagache, G.,
\&\ Hivon, E. 1997, Nature, in press (astro-ph/9706095)
\reference Kinney, A.L., Calzetti, D., Bica, E., \&\ Storchi-Bergmann,
T. 1994, \apj, 429, 172 (K94)
\reference Leitherer, C., \&\ Heckman, T.M. 1995, \apjs, 96, 9
\reference Madau, P. 1996, to appear in ``Star Formation Near and
Far'' (astro-ph/9612157)
\reference Madau, P., Ferguson, H.C., Dickinson, M.E., Giavalisco,
M., Steidel, C.C., \&\ Fruchter, A. 1996, \mnras, 283, 1388 (M96)
\reference Madau, P. 1997, to appear in ``ORIGINS'', ed. J.M. 
Shull, C.E. Woodward, and H. Thronson (astro-ph/9707141; M97)
\reference Meurer, G.R., Heckman, T.M., Leitherer, C., Kinney, A.,
Robert, C., \&\ Garnett D.R. 1995, \aj, 110, 2665 (paper-1)
\reference Meurer, G.R., Heckman, T.M., Lehnert, M.D., Leitherer,
C., \&\ Lowenthal, J.  1997, \aj, 114, 54 (paper-2)
\reference Mushotzky, R.F., \&\ Loewenstein, M. 1997, \apjl, 481, L63
\reference Pettini, M., Steidel, C.C., Dickinson, M., Kellogg, M.,
Giavalisco, M., \&\ Adelberger, K.L. 1997, to appear in 
``The Ultraviolet Universe at Low and High Redshift: Probing 
the Progress of Galaxy Evolution'', W.H. Waller, M.N. Fanelli, 
J.E. Hollis, \& A.C. Danks, eds. (astro-ph/9707200; P97)
\reference Puget, J.-L., Abergel, A., Bernard, J.-P., Boulanger, F.,
Burton, W.B., D\'esert, F.-X., \&\ Hartmann, D. 1996, \aap, 308, 5
\reference Rieke, G.H., Loken, K., Rieke, M.J., \&\ Tamblyn, P. 1993,
\apj, 412, 99
\reference Sawicki, M., \&\ Yee, H.K.C. 1997, \aj, submitted
\endreferences
}

\end{document}